\documentstyle[multicol,aps,psfig]{revtex}
\begin{document}

  \title{The Lowest Order Hadronic Contribution to the Muon $g-2$ Value
                              with Systematic Error Correlations}
  \author{D.H. Brown and W.A. Worstell}
  \address{Physics Department,
           Boston University,
           590 Commonwealth Avenue,
           Boston, MA 02215}
  \date{11 June 1996}  
  \maketitle
  \begin{abstract}

We have performed a new evaluation of the hadronic contribution to
$a_{\mu}=(g-2)/2$ of the muon with explicit correlations of systematic errors
among the experimental data on $\sigma( e^+e^- \to hadrons )$. Our result for
the lowest order hadronic vacuum polarization contribution is 
$ a_{\mu}^{had} = 702.6(7.8)(14.0) \times 10^{-10}$ 
where the the first error is statistical and the second is systematic. The
total systematic error contributions from below and above $\sqrt{s} = 1.4$ GeV
are $(13.1) \times 10^{-10}$ and $(5.1) \times 10^{-10}$ respectively, and are
hence dominated by the low energy region. Therefore, new measurements on
$\sigma( e^+e^- \to hadrons )$ below 1.4 GeV can significantly reduce the total
error on $a_{\mu}^{had}$. In particular, the effect on the total errors of new
hypothetical data with 3 \% statistical and 0.5 - 1.0 \% systematic errors is
presented. 

  \end{abstract}
  \pacs{}

\begin{multicols}{2}\narrowtext

\section{Role of Hadronic Contribution in $g-2$}

A new measurement of the anomalous magnetic moment of the muon, 
$ a_{\mu} \equiv (g-2)/2$, to an absolute accuracy of 
$\sigma_{exp}^{a_{\mu}} \sim \pm \ 4.0 \times 10^{-10}$
is proposed by the AGS E821 Collaboration at Brookhaven National Laboratory
\cite{E821} \cite{H92}. The theoretical value of the muon $g-2$ value consists
of at least the three standard model contributions: quantum electrodynamics
(QED), electroweak (EW) and hadronic. The latter arise from hadronic vacuum
polarization effects caused by effective photon couplings to hadrons via
charged quarks and consequent quantum chromodynamics (QCD) interactions with
gluons. 

Any residual difference between the sum of the standard model contributions and
the new experimental value $a_{\mu}^{exp}$ will be indicative of new physics: 
$$ a_{\mu}^{residual} = a_{\mu}^{exp} - a_{\mu}^{QED} - 
a_{\mu}^{EW} - a_{\mu}^{had} . $$
The new experimental value can only be sensitive
to electroweak and possibly supergravity \cite{SUGRA} \cite{CN95} 
and muon substructure effects \cite{MMPR90} provided
the errors on the standard model contributions are known 
better than the experimental accuracy. 
The QED and EW contributions have been calculated from theory
and are known an order of magnitude 
better than the expected experimental accuracy
($ a_{\mu}^{QED} = 11658470.6 \pm 0.2 \times 10^{-10}$ \cite{K95} and 
 $ a_{\mu}^{EW}  = 15.1 \pm 0.4  \times 10^{-10}$ \cite{CKM95}).
At the same time,
the lowest order hadronic contribution cannot be calculated 
accurately enough by QCD and hence a phenomenological procedure
must be used for its calculation. While the results of
the published $a_{\mu}^{had}$ calculations have errors which
vary from just above E821 error to greater than the electroweak
contribution (see Tab.~\ref{t:history})
and even differ in principle of approach,
it is of interest for the interpretation of a new
experimental measurement of the muon $g-2$ value
to investigate in more detail the 
precise procedure used for calculation 
of the hadronic contribution total error.

Fortunately the hadronic contribution to the muon $g-2$ value
can be related to a dispersion integral over
the experimental cross section $\sigma( e^+e^- \to hadrons )$
(see Sect.~\ref{s:bkgd}).
The error on the 
hadronic contribution to $a_{\mu}$ is hence determined by the error on the
experimental cross section and 
is by all acounts larger
than the expected E821 accuracy. Therefore, the hadronic contribution
error will largely determine the sensitivity level of E821 to new physics:
$$ \sigma_{residual} = \sqrt{ \sigma_{exp}^2 + \sigma_{QED}^2 + 
                              \sigma_{EW}^2 + \sigma_{had}^2 }
                     \sim \sqrt{ \sigma_{exp}^2 + \sigma_{had}^2 } .$$
This is the motivation for 
several recent evaluations of the lowest order hadronic
contribution to $a_{\mu}$ which are presented in 
Tab.~\ref{t:history}.
These calculations may roughly be placed into two 
categories of approach whether they
are based primarily upon (aggresive) model
dependent techniques which yield relatively smaller errors 
\cite{BINP85} \cite{DDS92} \cite{AY95} or on 
(conservative) model independent techniques (trapezoidal integration) 
which yield relatively larger errors \cite{KNO85} \cite{EJ95}.
A typical method for evaluating systematic error is comparison
of the model-dependent calculation with one based on trapezoidal integration.

While the model dependent calculations are based on theoretical innovations
to represent various components of 
the cross section $\sigma( e^+e^- \to hadrons )$
they lack the merit of a clear
prescription for correlating the systematic errors of the experimental
input. These errors are in some
cases as large as 20 \% in overall normalization of the cross section.
Therefore, the 
$\chi^2$ criterion used to select a set of fit parameters,
although indicative of the best overall fit to the data, 
may perhaps not
indicate the true error on the integral over experimental data points
which is the hadronic contribution to muon $g-2$. 
In particular, 
a shift of the whole curve up or down may fall within the experimental data
point errors and yet yield a variation in $a_{\mu}^{had}$ larger
than the stated errors.

The model independent techniques for calculating $a_{\mu}^{had}$
are based upon trapezoidal integration over the experimental data points.
As will be described below, this approach allows for a definite
procedure for determining both the central value for $a_{\mu}^{had}$ 
and its errors both of which make explicit use of the statistical and
systematic errors for each experimental data point input.
In particular, for each experiment an error weighted fraction 
will be defined and used in averaging over 
different experiments which make measurements of the same cross
sections over common energy regions.

The most recent model independent analysis \cite{EJ95} was the first to account
for systematic error correlations among the experimental data used
in the $a_{\mu}^{had}$ calculation. The systematic errors
were correlated at each energy point 
separately in the process of determining an error-weighted average
cross section ratio $R(s)$ (see Sect.~\ref{s:bkgd})
{\it before} performing the $g-2$ integral.
Special care was taken to calculate the $\rho$-meson contribution
separately from the rest of the hadronic production cross sections
and there are many significant contributions to the 
discussion of $a_{\mu}^{had}$ calculations
in this important communication.

The main new aspect of the present $a_{\mu}^{had}$ 
calculation 
is that the systematic errors are correlated 
by a different error combination formula
{\it after} performing the $g-2$ integral.
Furthermore, correlations are accounted for across energy regions and
between experimental measurements of the $\rho$-meson and of the 
additional hadronic cross sections which contribute
to $\sigma( e^+e^- \to hadrons )$. (It appears that
by treating the $\rho$-meson
separately the previous calculation did not correlate some
errors completely.)

In short, 
the $g-2$ integral has been performed over each hadronic
cross section separately without combining them into their equivalent
$R(s)$ value
for the present calculation.
The focus is therefore upon 
the error-weighted average of the
$a_{\mu}^{had}$ integral value itself (and its errors).
These innovations have yielded a result with total error
remarkably equal to that obtained previously \cite{EJ95}.
However,
there is a difference obtained as to how much of the error
derives from the energy region above and below 1.4 GeV
which calls attention to the precise procedure of error combination
employed in the calculation.

\section{Background on \hbox{$a_{\mu}^{had}$}\ Calculations}
\label{s:bkgd}

\begin{figure}
\hbox{ \centerline{
\psfig{file=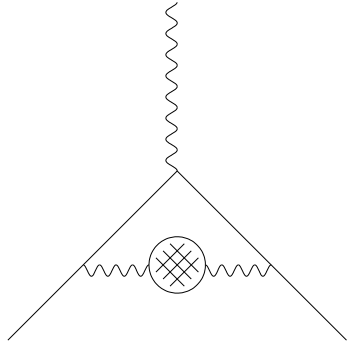,height=5cm,width=5cm,rheight=5cm,rwidth=5cm}
} }
\vspace{0.25in}
\caption{Lowest order hadronic contribution to \hbox{$a_{\mu}^{had}$}.}
\label{f:had1}
\end{figure}

\begin{figure}
\hbox{ 
\psfig{file=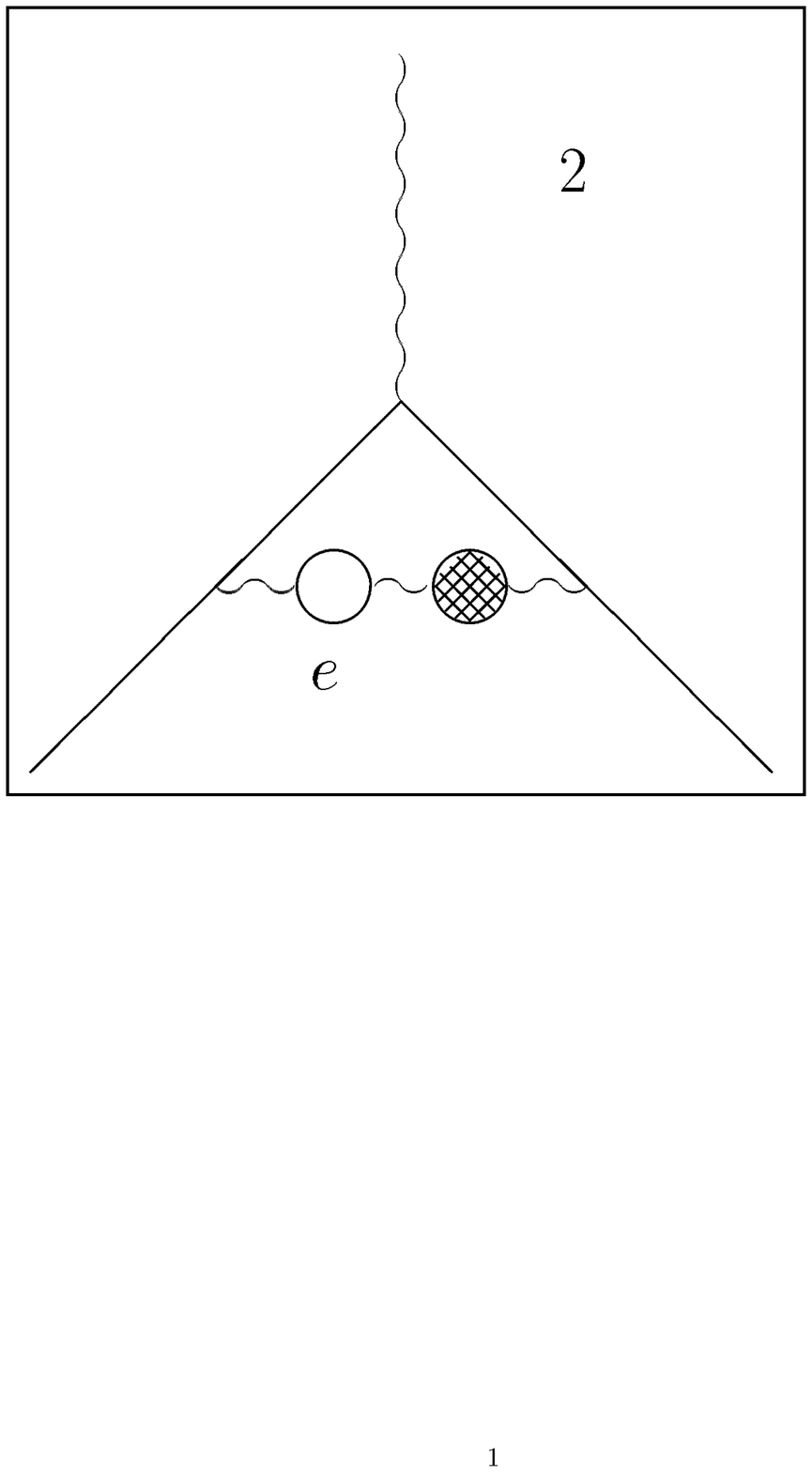,height=10cm,width=8cm,rheight=5cm,rwidth=2cm}
\hskip 1.0in
\psfig{file=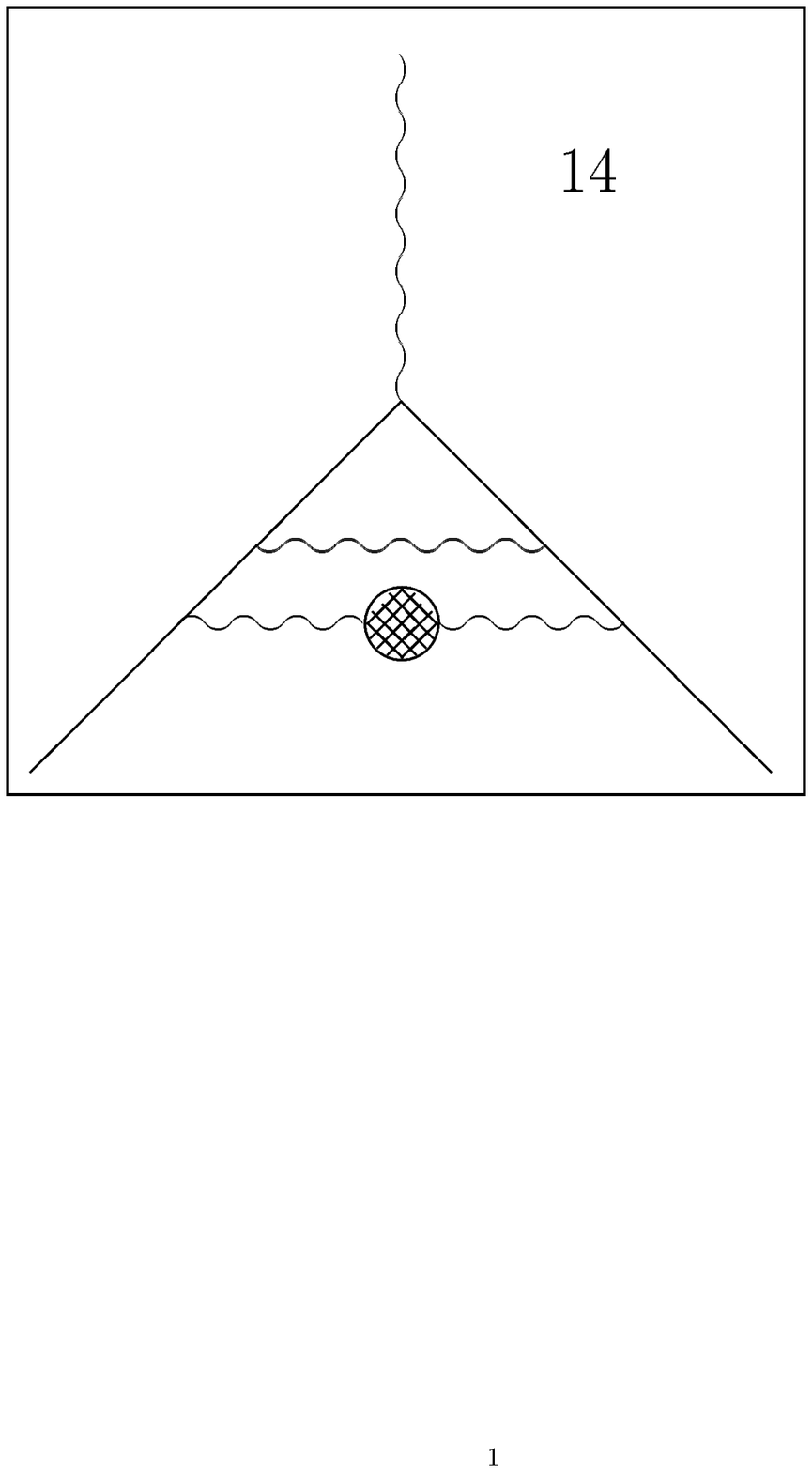,height=10cm,width=8cm,rheight=5cm,rwidth=2cm}
\hskip 1.0in
\psfig{file=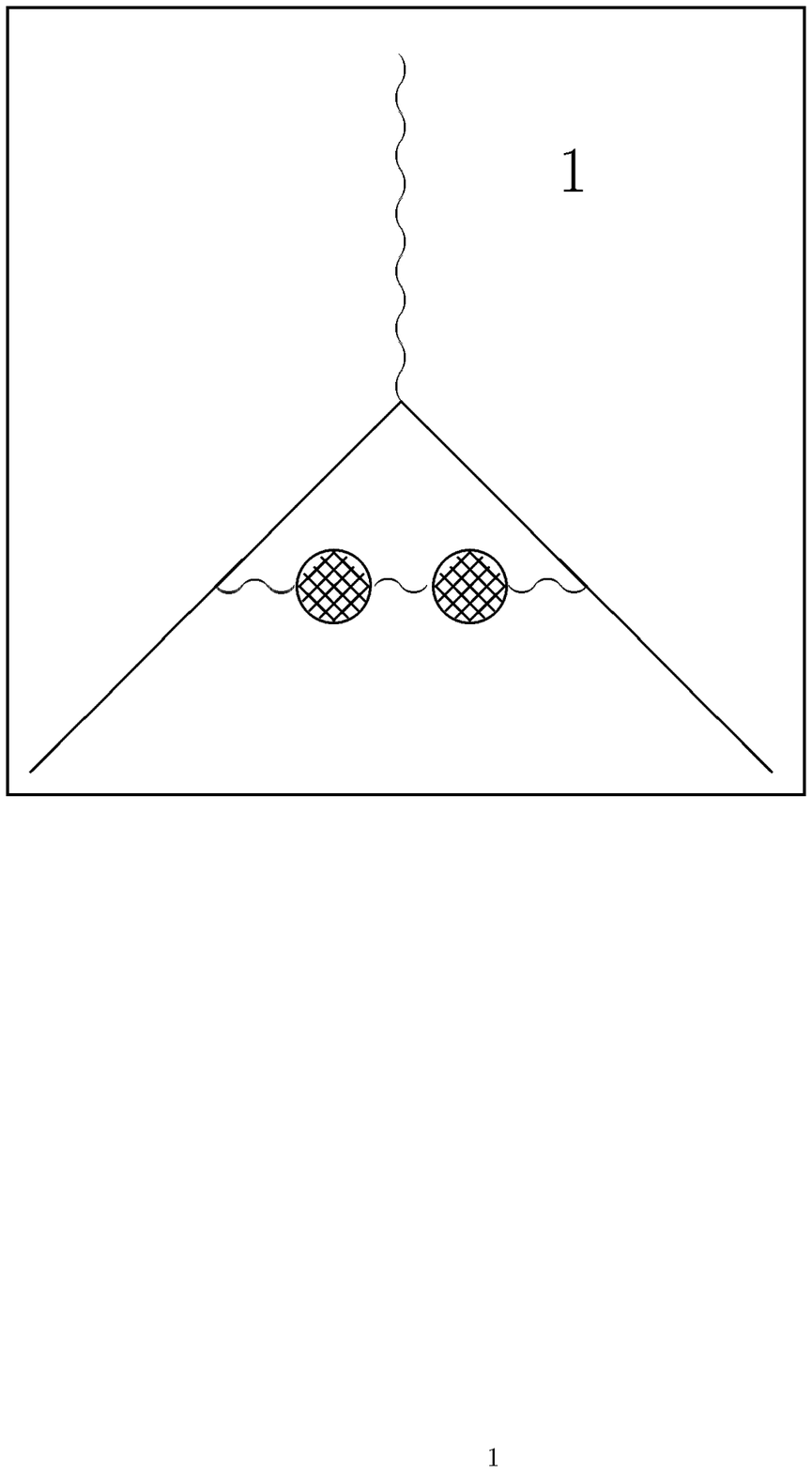,height=10cm,width=8cm,rheight=5cm,rwidth=2cm}
} 
\vspace{0.5in}
\caption{Higher order hadronic contributions to \hbox{$a_{\mu}^{had}$}}
\label{f:had2}
\end{figure}

The hadronic contribution to \hbox{$a_{\mu}$}\ consists of 
a dominant lowest order term,
shown in Fig.~\ref{f:had1}, several higher order terms 
in Fig.~\ref{f:had2} (the number above each
diagram indicates how many
contributing diagrams are in its class), and finally a group
of hadronic light-by-light scattering terms.
Detailed calculations are given elsewhere of the 
Figure~\ref{f:had2} and hadronic light-by-light
contributions. Since their errors are 
well below the expected E821 error of
$ \sigma_{exp}^{a_{\mu}} \sim \pm \ 4.0 \times 10^{-10}$
($ a_{\mu}^{Fig2} = -9.0 \pm 0.5 \times 10^{-10}$ \cite{KNO85} and 
 $ a_{\mu}^{light} = -5.2 \pm 1.8 \times 10^{-10}$ \cite{HKS95}) 
they are not discussed here.
There is another hadronic light-by-light calculation with a larger
error $ a_{\mu}^{light} = -9.2 \pm 3.2 \times 10^{-10}$ \cite{BPP96} 
but it does not depend
on the cross section $\sigma(e^+ e^- \to hadrons)$ and therefore
is not of concern to the present calculation.
In view of the difference between the two
latest hadronic light-by-light contributions, and the consequent
ambiguity over defining the {\it total} hadronic contribution,
this paper is concerned only with
the lowest order hadronic contribution 
to the muon $g-2$ value
shown in Fig.~\ref{f:had1}.

\subsection{Formalism of the Hadronic Contribution \hbox{$a_{\mu}^{had}$}}
\label{s:formalism}

The largest contribution to \hbox{$a_{\mu}^{had}$},
shown in Fig.\ \ref{f:had1}, can be related to 
the total Born cross section (lowest order in QED)
for hadron production in electron-positron annihilations,
$\sigma^0_{had} = \sigma^0 ( e^+ e^- \to hadrons )$,
by means of dispersion theory and the optical
theorem \cite{LPR72}. Defining $ \xi \equiv s / m_{\mu}^2 $
and $\beta \equiv \sqrt{ 1 - 4 / \xi }$ the result is

\end{multicols}

\widetext

\begin{equation}
\hbox{$a_{\mu}^{had}$} = { 1 \over 4 \pi^3 } 
\int_{4 m_{\pi}^2}^{\infty} {\rm d}s \ 
\sigma^0_{had} \ K (s)  =
{ m_{\mu}^2 \over 9 }
\left( \alpha \over \pi \right)^2
\int_{4 m_{\pi}^2}^{\infty} {\rm d}s \ 
R(s) \ K_2 (s) \ { 1 \over s^2 }
\label{e:amuhad}
\end{equation}
where the kernel function $K(s)$ in general arises from a massive
photon propagator in the $g-2$ Schwinger calculation:
\begin{displaymath}
  K (s) = \int_0^1 {\rm d}x
{ x^2 \left( 1 - x \right) \over x^2 + \left( 1 - x \right) \xi } 
       \to_{s \to \infty} { m_{\mu}^2 \over 3 s }
\end{displaymath}
\begin{equation}
K(s) = 
{ 1 \over 2 \beta } 
  \left[
{ \left( 1 - \beta \right)^2 \over 1 + \beta }
  \left( 1 - { 2 \over 1 + \beta } \ln { 2 \over 1 - \beta } \right) -
{ \left( 1 + \beta \right)^2 \over 1 - \beta }
  \left( 1 - { 2 \over 1 - \beta } \ln { 2 \over 1 + \beta } \right)
                                                             \right] -
{1 \over 2} .
\label{e:ourK}
\end{equation}

\begin{multicols}{2}\narrowtext

The cross section ratio $R(s) \equiv \sigma^0_{had} / \sigma^0_{\mu\mu}$
with $\sigma^0_{\mu\mu}= 4 \pi \alpha^2 / 3 s$ and
the kernel $K_2(s)$ is used in the $R(s)$ formulation of the
$a_{\mu}^{had}$ dispersion integral:
\begin{displaymath}
K_2 (s) \equiv { 3 s \over m_{\mu}^2 } K (s).
\end{displaymath}

The $\sigma_{had}$ formulation is useful for
low energy data which are usually published as individual
exclusive hadronic cross sections;
the $R(s)$ form of the $a_{\mu}^{had}$ 
dispersion integral is useful for higher
energies where experimental data are usually published as the inclusive
ratio $R$.

From the $a_{\mu}^{had}$ dispersion integrals it is apparent that
the error on the hadronic cross sections determines
the error on $a_{\mu}^{had}$. Therefore,
the AGS E821 experimental error goal just larger than $3.5 \times 10^{-10}$ 
on the hadronic contribution of roughly $700 \times 10^{-10}$ requires
a \hbox{0.5 \%} accuracy on the hadronic contribution calculation.

\section{The WFSA Evaluation Procedure}
\label{s:our}

\subsection{Correlation Postulates}
\label{s:post}

As is implied by the $g-2$ dispersion integral (Eq.~\ref{e:amuhad})
the $a_{\mu}^{had}$ calculation procedure
consists of some combination of the following steps:
1) integration over energy, 2) weighted average over detectors
and 3) sum over exclusive hadronic modes which contribute
to $\sigma^0 (e^+e^- \to hadrons)$.
The sequence of these three sums may be interchanged
as for example performing the energy integration last \cite{EJ95}.
Most calculations, however, employ the sequence (1,2,3) $-$ the
most natural one $-$ where in the last step, the quadrature sum of errors
over modes implies the assumption of zero correlation of systematic
errors among experiments (a reasonable first approximation).

To go beyond the first approximation, it is necessary to survey the
existing experimental data on hadronic production cross sections.
The published data and recent preprints on exclusive hadron production in
electron-positron annihilations used in the present calculation are listed
in Appendix~\ref{a:data} where it is evident that eight detectors
have measured more than one mode. Since a given detector uses the same
luminosity and similar correction factor calculations (e.g. radiative
corrections, efficiencies which use some of the same subroutines)
it is reasonable to suppose that the cross section determinations
of different exclusive hadronic modes by a single detector 
may in fact be correlated.
The following correlation postulates are therefore intended
to address this situation:

\begin{enumerate}

\item
A single detector measuring more than one exclusive hadronic mode
has 100 \% correlations among systematic errors due to common
luminosity and correction factor calculations;

\item
Different detectors have uncorrelated errors since they do not
share luminosity and correction factor calculations.

\end{enumerate}

The accomodation of these correlation postulates requires
a particular sequence of the three sums (1,2,3) 
for combination of the $a_{\mu}^{had}$ central values
and an alternative sequence (1,3,2) for the $a_{\mu}^{had}$ errors.
In both cases,
the energy is integrated over first and separately for each
detector and exclusive hadronic mode measured
over energy sub-regions, where these sub-regions are
defined by common energy coverage among detectors.

The former sequence (1,2,3) facilitates the need to first calculate
the error weighted fractions (defined in detail below) while the latter
sequence (1,3,2) is necessary for, according to postulate 1,
correlating individual detector systematic errors over the modes
measured by that detector. This must be done before the final
uncorrelated combination of errors is made, according to postulate 2,
across detectors.
As this method is based on {\sl W}eighted {\sl F}raction
averaging with {\sl S}-factor application (see Eq.~\ref{e:S})
{\sl A}fter $a_{\mu}^{had}$
integration over the energy, it is here referred to as the
WFSA method.

Lastly, in the present calculation it is noted that
exclusive hadronic cross sections (modes)
up to 2.0 GeV have been used because this can reveal the 
propagation of errors from each detector and exclusive mode separately.
An additional consideration, although less important, is that
the exclusive hadronic mode spectra
may contain interference effects (when more than one vector meson
contributes to a given exclusive hadronic mode) which would 
otherwise require special care if individual vector meson contributions
were calculated separately.
Further, the uncertainty over
the generation mechanism, between the $e^+e^-$ annihilation photon
and the hadronic final state, is avoided by focusing on exclusive hadronic
modes themselves.

For the energy region 2.0 to 3.1 GeV, the inclusive cross section ratio
$R(s)$ has been used (in the absence of exclusive data in this energy region)
with the WFSA procedure for the contributing experiments.
In the region above 3.1 GeV, the QCD expression has been used without
the WFSA procedure since perturbative QCD is expected to be valid
(see Sect.~\ref{s:QCD}).

\subsection{Trapezoidal Integration Procedure}

The usual trapezoidal integration technique takes
the experimental data points pairwise: the cross sections, 
systematic and squared statistical errors are averaged per pair, 
then multiplied by the energy width of the pair, and finally they are
summed over all pairs. However, it is convenient for treatment of
statistical errors to expand the sum in order to remove terms which cancel.
Denoting by $s_k$, $K_k$, $c_{ijk}$, 
$\sigma_{ijk}^{stat}$, and $\sigma_{ijk}^{sys}$,  
the energy, kernel function, the cross section
and its statistical and systematic error
from the $i$th detector, $j$th
exclusive mode, at the $k$th energy point,
the integration of Eq.\ (\ref{e:amuhad}) can be represented by:
\begin{eqnarray*}
a_{ij} &=& { 1 \over 4 \pi^3 } \ \ \sum_{k=1}^{n-1}
\left\{ {1 \over 2} \left( c_{ijk} K_k + c_{ij,k+1} K_{k+1} \right) \right\}
                                 \left( s_{k+1} - s_k \right) \\
       &=& { 1 \over 4 \pi^3 } \ {1 \over 2} \ 
\left( A_1 + \sum_{k=2}^{n-1} A_k + A_n \right) \\[4mm]
A_1 &=& c_{ij,1} K_1 ( s_2 - s_1 )         \\ 
A_k &=& c_{ij,k} K_k ( s_{k+1} - s_{k-1} ) \\
A_n &=& c_{ij,n} K_n ( s_n - s_{n-1} )
\end{eqnarray*}
where the first and last terms in the sum are handled separately
and the middle terms have an energy width across both upper and lower
neighboring data points instead
of across the points in pairs. This latter form is 
necessary for the proper treatment of the statistical errors:
\begin{eqnarray*}
\sigma_{ij}^{stat} &=& { 1 \over 4 \pi^3 } \ { 1 \over 2 } \ 
\sqrt{ {\sigma_1}^2 + \sum_{k=2}^{n-1} {\sigma_k}^2 + {\sigma_n}^2 } \\[4mm]
{\sigma_1}^2 &=& (\sigma_{ij,1}^{stat} K_1)^2 \left( s_{2}-s_{1} \right)^2  \\
{\sigma_k}^2 &=& 
        (\sigma_{ij,k}^{stat} K_k)^2 \left( s_{k+1} - s_{k-1} \right)^2  \\  
{\sigma_n}^2 &=& (\sigma_{ij,n}^{stat} K_n)^2 \left( s_{n} - s_{n-1} \right)^2.
\end{eqnarray*}

The systematic errors are specified by an array of values 
$p_{ijk}^{sys}$ which are given as a percentage of the total cross section.
Since the \hbox{$a_{\mu}^{had}$}\ contribution terms above 
are linear in the cross section
this implies that the systematic errors should be:
\begin{displaymath}
 \sigma_{ij}^{sys}  = 
                 p_{ij,1}^{sys} A_1 +  
\sum_{k=2}^{n-1} p_{ij,k}^{sys} A_k +  
                 p_{ij,n}^{sys} A_n    \nonumber .
\end{displaymath}
Usually the systematic error is the same for all energy points; 
however, it is different for each point of
the NA7 and Olya \hbox{$\pi^+ \pi^-$}\ measurements.
In the cases where
no systematic errors were given or readily located in the literature,
values from comparable measurements with the same detector have been taken
wherever possible, or as 10 \% if the statistical errors dominate.
In addition, a calculation with 20 \% for these
ambiguous values has shown
that 
the WFSA results do not depend on the arbitrary selection of 10 \% 
\cite{Note220}.
All of the systematic errors that were used are listed in Tab.\ \ref{t:systpp}
for \hbox{$\pi^+ \pi^-$}, Tab.\ \ref{t:systppp} for \hbox{$\pi^+ \pi^- \pi^0$}, 
Tab.\ \ref{t:systothr} for the higher multiplicity modes, and
Tab.\ \ref{t:syst23} for the energy region 2.0 - 3.1 GeV.

If the desired limits of integration are inside (outside) 
the given data energy range,
the cross section and error are linearly interpolated 
(extrapolated) to give the relevant
pair of points.
Thus our energy ranges are variable 
and have been set to match each of the previous evaluations for detailed
comparisons \cite{Note220}.

\subsection{Weighted Fraction Averaging}
\label{s:WFSA}

In order to arrive at an \hbox{$a_{\mu}^{had}$}\ 
contribution per exclusive mode,
\hbox{$a_{\mu}^{mode} \equiv a_j$}, 
error-weighted averages across detectors have been performed according to
the PDG\cite{PDG94} procedure (for the $i$th detector and $j$th exclusive
hadronic mode): 
\begin{equation}
a_j \pm \sigma_j = 
{ \sum_i w_{ij} a_{ij} \over \sum_i w_{ij} } \pm 
\left( 1 \over \sum_i w_{ij} \right)^{1/2}
\label{e:WF}
\end{equation}
where
\begin{equation}
w_{ij} = { 1 \over \sigma_{ij}^2 } \ ; \ 
\sigma_{ij}^2 = \left( \sigma_{ij}^{stat} \right)^2 +
                \left( \sigma_{ij}^{sys } \right)^2 .
\label{e:error}
\end{equation}
In addition, the quality of the error weighted
combinations were assessed by calculating the $\chi^2$ and PDG scale 
factor (i.e. $\chi^2$ per degree of freedom) \cite{PDG94}:
$$ \chi^2_j = \sum_i^N w_{ij} \left( a_j - a_{ij} \right)^2 $$
\begin{equation}
S_j \equiv \sqrt{ \chi^2_j \over N - 1 } \label{e:S}
\end{equation}
where $N$ is the number of detectors included in the average.
If $S>1$ then the errors were 
scaled up by this factor.

At this point the prescription 
for determining the error-weighted fractional contribution to the total
error from a given statistical or systematic error
of the $i$th detector and $j$th exclusive mode is needed.
For this purpose
the PDG expression above for total
squared error on a weighted average (see Eq.~\ref{e:WF}) 
can be expanded as a sum over
squared component errors where in the first step
the trivial sum $\sum_{i=1}^N = N$ is used to multiply by unity:
\begin{eqnarray*}
{\sigma_j}^2
 = {1 \over \sum_{i} w_{ij} } 
&=& {1 \over \sum_{i} w_{ij} } 
\left( {1 \over N} \sum_{i=1}^N { \sigma_{ij}^2 \over \sigma_{ij}^2 } 
\right) \\
&=& \sum_{i=1}^N {1 \over N} { w_{ij} \over \sum_{i} w_{ij} } 
\sigma_{ij}^2 .
\end{eqnarray*}
The last step makes use of the definition 
$ w_{ij} = 1 / \sigma_{ij}^2 $ 
while the definition for the remaining $\sigma_{ij}^2$ in the numerator
(see Eq.~\ref{e:error})
leads to the separation of squared
statistical from systematic terms in the sum:
\begin{eqnarray}
{\sigma_j}^2
&=& \sum_{i=1}^N \left[
\left( \overline{ \sigma_{ij}^{stat} } \right)^2 +
\left( \overline{ \sigma_{ij}^{ sys} } \right)^2          
\right] 
\nonumber
\\
\overline{ \sigma_{ij}^{stat} } 
&=& 
\sqrt{
{1 \over N} { w_{ij} \over \sum_{i} w_{ij} } 
}
\ S \ \sigma_{ij}^{stat}  
\label{e:WFstat} \\
\overline{ \sigma_{ij}^{sys} } 
&=& 
\sqrt{
{1 \over N} { w_{ij} \over \sum_{i} w_{ij} } 
}
\ S \ \sigma_{ij}^{sys} .
\label{e:WFsys}
\end{eqnarray}
Note that the PDG scale factor $S$ has been inserted to
emphasize the fact that the errors are to be increased
if and only if the scale factor $S > 1$.
(These expressions differ from those in the previous
$a_{\mu}^{had}$ calculation \cite{EJ95} by the factors $1 / N$ and $S$.)

Armed with these expressions it is possible to 
implement the correlation postulates
for the $i$th detector by
summing the {\sl averaged} systematic errors
(i.e. the weighted fractional systematic error
contributions to the total errors in Eq.~\ref{e:WFsys})
linearly over contributing exclusive modes,
while leaving weighted fractional statistical errors (Eq.~\ref{e:WFstat})
uncorrelated.
The correlation postulates are simply executed by the following sums:
\begin{eqnarray}
\overline{ \sigma_{i}^{sys} } 
&=&
\sum_j 
\overline{ \sigma_{ij}^{sys} } \\
\overline{ \sigma_{i}^{stat} } 
&=& 
\sqrt{
\sum_j 
\left( \overline{ \sigma_{ij}^{stat} } \right)^2
} .
\end{eqnarray}
Note that $\sigma_j$ itself is not directly used to determine the
total WFSA error; to sum $\sigma_j$ over $j$ in quadrature
would yield a total
error which ignores correlations. This is what most previous
$a_{\mu}^{had}$ calculations have done.

\section{Application of the WFSA Method}

\subsection{Correlations Over Energies}

In general, if different detectors measure hadronic cross sections
over different energy regions, there must be different error-weighted
fractions and $S$-factors for each common energy region
($i$th detector, $j$th energy region).
Such a WFSA application over energies
has been applied to the
dominant 
\hbox{$a_{\mu}^{had}$} 
contributions from the
hadronic modes: \hbox{$\pi^+ \pi^-$}, 
\hbox{$\pi^+ \pi^- \pi^0$}, \hbox{$K^+ K^-$}, and \hbox{$K_L K_S$}.
An example of WFSA correlations over energy is
shown in Tab.~\ref{t:Rwf} for 
the $R(s)$ 
\hbox{$a_{\mu}^{had}$} 
contribution in the energy range from 2.0 to 3.1 GeV.
The $a_{\mu}^{had}$ central values are error-weighted
averaged across detectors
(horizontally in the table) and summed over energies (vertically)
while the detector specific statistical (systematic) errors are combined 
in quadrature (linearly) across the energies (vertically).
The correlation of systematic errors over energies does not
appear to be taken into account by the WFB method discussed in the
previous $a_{\mu}^{had}$ calculation \cite{EJ95}: 
{\it W}eighted {\it F}raction averaging (of $R(s)$ values) {\it Before} energy 
integration. 

In this energy range,
published inclusive
$R(s)$ values have been used
from the \hbox{$\gamma \gamma 2$} detector \cite{GG279} \cite{GG281}
and MARK I \cite{MarkI}.
It is noted that the 
\hbox{$\gamma \gamma 2$} collaboration has published
values for $R_2(s)$ (two hadrons exclusively) \cite{GG279} and 
$R_{\geq 3}(s)$ (three or more hadrons) \cite{GG281}
separately which are added for the present calculation.
This does not appear to have been done in the
previous 
$a_{\mu}^{had}$ calculations.
In addition,
data published as $\sigma(e^+e^- \to hadrons)$ by the BCF 
collaboration
\cite{BCF74}
(not apparently included previously)
has been divided 
by $\sigma^0_{\mu\mu}$ to make them also $R(s)$ values.

\subsection{Correlations Over Modes}

If the hadronic cross sections are measured
over similar energy regions, as in the case of the 
$>$ 2 hadrons multiplicity cross sections shown in
Tab.~\ref{t:hmm},
then the WFSA method can simply be
applied over the modes ($i$th detector, $j$th hadronic mode).
In this case, a single error-weighted fraction and $S$-factor
suffices for evaluation of $a_{\mu}^{had}$ central value and errors.
As before,
the $a_{\mu}^{had}$ central values are error-weighted
averaged across detectors
(horizontally in the table) and summed over modes (vertically),
while the detector specific statistical (systematic) errors are combined 
in quadrature (linearly) across, in this case, 
the modes (vertically). The errors are not to
be combined across detectors until all modes have been treated separately
and then taken together.
It is this correlation over all modes, in particular correlating 
the $\pi^+ \pi^-$ detector total systematic errors
with all the 
other two body and higher multiplicity modes, over all energies,
which appears not to have been
done before. (The previous $a_{\mu}^{had}$ calculation \cite{EJ95}
appears to have accounted for these correlations only above 0.81 GeV,
after the peak of the $\rho$-meson. Hence, it appears that 
the $\pi^+\pi^-$ systematic
errors below 0.81 GeV were not correlated
with systematic errors of measurements 
by the same detectors of the other modes over all energies and
of the $\pi^+ \pi^-$ mode above 0.81 GeV.)

\subsection{Kaons, Narrow Resonances and Higher VMD modes}
\label{s:kaons}

Some additional hadronic
modes which contribute to the total hadronic production cross section
$\sigma^0 (e^+ e^- \to hadrons)$
require comment.
In particular, experimental data in the form of total
cross sections neither
exists on the radiative decays
of the $\omega$ and $\phi$ mesons ($\pi^0 \gamma, \eta \gamma$)
nor on the kaon pair production
of the $\phi$ meson below certain energies.
Further, 
there are additional contributions to hadronic vacuum 
polarization than from 
just the lowest order single vector meson dominated (VMD)
amplitudes represented by the hadronic (decay) modes previously discussed.

\subsubsection{Kaons and Narrow Resonances}

Experimental data on the total cross sections for
production of kaon pairs (charged and neutral) are limited by the fact
that nuclear interactions of low momentum kaons are not
well measured. Hence kaon detection efficiencies are difficult to
calculate precisely, and although existing data does span both sides of the
$\phi$ meson it is usually a {\it relative} cross section useful
for measurements of $\phi$ meson parameters. Hence 
a Breit-Wigner (BW) line shape with PDG 1994 parameters
has been used, below the (most recent) lowest
data point for kaon pair production and for the whole energy range
for the radiative decays, where the errors (considered totally systematic)
are evaluated by differentiating the BW formula.

In the charm and bottom threshold regions
the 6 states each of the $J/\Psi$ 
and $\Upsilon$ resonance families have been calculated
separately using
PDG 1994 values and the peak approximation formula for $a_{\mu}^{had}$
(\cite{BLO75} \cite{KNO85}). In view of the small contribution
to $a_{\mu}^{had}$ from the $b$-quark (due to kernel function
supression in Eq.~\ref{e:amuhad}), the top quark contribution
is neglected.

\subsubsection{Higher Order VMD Contributions}

\begin{figure}
\psfig{file=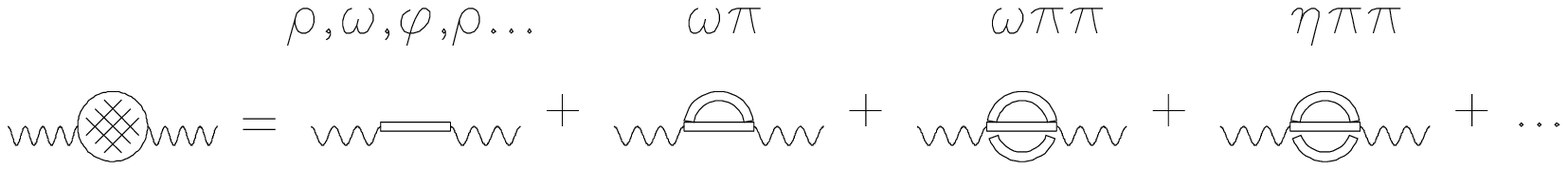,height=7cm,width=7cm,rheight=7cm,rwidth=7cm}
\caption{Vector Meson Dominance representation of hadronic
vacuum polarization.}
\label{f:VMD}
\end{figure}

The Vector Meson Dominance (VMD) model approximation of QCD
expresses the fact that vector mesons, instead of quarks and gluons,
are the relevant degrees of freedom in low
energy (near threshold) QCD interactions. The VMD model approximation
for hadronic vacuum polarization is depicted in Fig.~\ref{f:VMD}.
Most of the $a_{\mu}^{had}$ contributions are contained in the first
term on the right of the figure. The decay modes of the vector mesons $\rho$,
$\omega$, $\phi$ and $\omega'$ account for 
$\sim 597 \times 10^{-10}$ (85\% of $a_{\mu}^{had}$), while
$\rho'$ accounts for $\sim 34.7 \times 10^{-10}$ (5\% of $a_{\mu}^{had}$),
and $\phi'$ for $\sim 1.6 \times 10^{-10}$ by rough accounting.
The remaining part ($\sim 10\%$) of $a_{\mu}^{had}$ 
is derived from a non- or very broadly resonant background
for which it is instructive to consider some additional
generation mechanisms not already included in the previous exclusive
mode approach.

Namely,
some portion of $a_{\mu}^{had}$ derives from 
higher order VMD interactions $V\pi$ and $V\pi\pi$ (the next terms
on the right in Fig.~\ref{f:VMD})
where $V$ is
a vector meson. Although these amounts are individually less than the
expected E821 error ($\sim 4.0 \times 10^{-10}$) the fact that they
may combine with
other small contributions (whose sum may be greater than the
E821 error) implies that they all should indeed be carefully considered.

The cross sections
$\sigma(\hbox{$K^* K^{\pm} \pi^{\mp}$})$, 
$\sigma(\omega \pi^0)$, 
$\sigma(\omega \hbox{$\pi^+ \pi^-$})$, 
$\sigma(\eta \hbox{$\pi^+ \pi^-$})$
have recently (1991,1992) been measured (ND,DM2).
In particular, it has been 
pointed out \cite{EJ95} 
that the following exclusive hadronic modes 
marked with the comment ``\underline{not}" need  
to be included (B refers to the branching fraction for the decay mode
in parentheses): 
\bigskip

\begin{tabular}{rclll}
$\omega \pi^0$
&$\to$& $(\hbox{$\pi^+ \pi^- \pi^0$}) \pi^0$
                  & B=0.888  & is in \hbox{$\pi^+ \pi^- \pi^0 \pi^0$}  \\
&$\to$& $(\pi^0 \gamma) \pi^0$
                  & B=0.085  & \underline{not} counted    \\
&$\to$& $(\hbox{$\pi^+ \pi^-$}) \pi^0$
                  & B=0.0221 & is in \hbox{$\pi^+ \pi^- \pi^0$}     \\[2mm]
$\omega \hbox{$\pi^+ \pi^-$}$
&$\to$& $(\hbox{$\pi^+ \pi^- \pi^0$}) \hbox{$\pi^+ \pi^-$}$
                  & B=0.888  & is in \hbox{$\pi^+ \pi^- \pi^+ \pi^- \pi^0$}  \\
&$\to$& $(\pi^0 \gamma) \hbox{$\pi^+ \pi^-$}$
                  & B=0.085  & \underline{not} in \hbox{$\pi^+ \pi^- \pi^0$} \\
&$\to$& $(\hbox{$\pi^+ \pi^-$}) \hbox{$\pi^+ \pi^-$}$
                  & B=0.0221 & is in \hbox{$\pi^+ \pi^- \pi^+ \pi^-$} \\[2mm]
$\omega \pi^0 \pi^0$
&$\to$& $(\hbox{$\pi^+ \pi^- \pi^0$}) \pi^0 \pi^0$
                  & B=0.888  & is in \hbox{$\pi^+ \pi^- \pi^0 \pi^0 \pi^0$}  \\
&$\to$& $(\pi^0 \gamma) \pi^0 \pi^0$
                  & B=0.085  & \underline{not} counted \\
&$\to$& $(\hbox{$\pi^+ \pi^-$}) \pi^0 \pi^0$
                  & B=0.0221 & is in \hbox{$\pi^+ \pi^- \pi^0 \pi^0$}.
\end{tabular}

\noindent
They further point out that isospin considerations \cite{Pais} imply
\begin{eqnarray*}
2 \sigma( \hbox{$\pi^+ \pi^- \pi^0 \pi^0 \pi^0$} )
&=& \sigma( \hbox{$\pi^+ \pi^- \pi^+ \pi^- \pi^0$} ) \\
2 \sigma( \omega \pi^0 \pi^0 )  &=& \sigma( \omega \hbox{$\pi^+ \pi^-$} )
\end{eqnarray*}
and that this must be used since $\sigma( \omega \pi^0 \pi^0 )$
and $\sigma( \hbox{$\pi^+ \pi^- \pi^0 \pi^0 \pi^0$} )$ have not been measured.
Therefore the $\sigma ( \pi^+ \pi^- \pi^+ \pi^- \pi^0 )$ contribution
has been augmented by a factor $A_1=1.5$,
and the contributions from $\sigma ( \omega \pi^0 )$ and
$\sigma ( \omega \hbox{$\pi^+ \pi^-$} )$ will contribute with
factors $B_1 = 0.085$ and $B_3 = 1.5 B_1$ respectively,
as noted in Tab.\ \ref{t:hmm}.
However, the same logic applied to
$\sigma( \eta \hbox{$\pi^+ \pi^-$})$
implies {\it not} including 100\% of it as done previously \cite{EJ95}:

\begin{tabular}{rclll}
$\eta \hbox{$\pi^+ \pi^-$}$
&$\to$& $(\gamma \gamma) \hbox{$\pi^+ \pi^-$}$
& B=0.388  & \underline{not} in \hbox{$\pi^+ \pi^-$}     \\
&$\to$& $(\pi^0 \pi^0 \pi^0) \hbox{$\pi^+ \pi^-$}$
& B=0.319  & is in \hbox{$\pi^+ \pi^- \pi^0 \pi^0 \pi^0$} \\
&$\to$& $(\hbox{$\pi^+ \pi^- \pi^0$}) \hbox{$\pi^+ \pi^-$}$
& B=0.236  & is in \hbox{$\pi^+ \pi^- \pi^+ \pi^- \pi^0$}  \\
&$\to$& $(\hbox{$\pi^+ \pi^-$} \gamma) \hbox{$\pi^+ \pi^-$}$
& B=0.0488 & \underline{not} in \hbox{$\pi^+ \pi^- \pi^+ \pi^- \pi^0$}.
\end{tabular}

\noindent
To avoid double counting in the \hbox{$\pi^+ \pi^- \pi^+ \pi^- \pi^0$}\
and \hbox{$\pi^+ \pi^- \pi^0 \pi^0 \pi^0$}\ channels
only the fraction $B_2 = 0.388 + 0.0488$ of the
$\sigma( \eta \hbox{$\pi^+ \pi^-$})$
contribution has been included. 
(There is no further factor of 1/2 for inclusion of
the cross section $\sigma( e^+e^- \to \eta \pi^0 \pi^0)$ since it is forbidden.)
The augmentation and branching factors ($A_1$, $B_1$, $B_2$, $B_3$)
are noted in Tab.\ \ref{t:hmm} where all contributions are listed
in descending order.

\subsection{The Perturbative QCD Energy Region}
\label{s:QCD}

To test the validity of QCD for determination of the 
contribution to $a_{\mu}^{had}$
in the energy region above 3.1 GeV,
a QCD parameterization 
\cite{Marshall} 
(including second order terms)
was compared with
a data-based evaluation \cite{EJ95}.
An asymptotic kernel function \cite{EJ95} was used
in the $a_{\mu}^{had}$ integral
since Eq.~\ref{e:ourK} is numerically stable only up to $\sim$ 20 GeV.
The errors on the QCD contribution to $a_{\mu}^{had}$
were determined by its variation with 
$\Lambda_{\overline{MS}} \pm \Delta \Lambda_{\overline{MS}}$.

Marshall combined the results from 15 different
$e^+e^-$ annihilation
experiments, fitting them to a third order QCD model with a
single parameter, $\Lambda_{\overline{MS}}$.  Because the fit was
overconstrained, he was able to evaluate whether these
experiments had overestimated their systematic errors.
He then went a step further, and fitted for the absolute
normalization of each of the 15 experiments independently,
bounded by double their stated systematic errors ($\pm 2 \sigma$).
He found fitted normalizations for most experiments within 
their stated limits, with the exception of two of the earliest 
experiments: Mark II and $\gamma \gamma 2$. 

By using the error
in his normalization constants (from the fit) rather than the
stated systematic errors for each experiment, Marshall was
able to significantly reduce his error on $\Lambda_{\overline{MS}}$ and
hence on his overall normalization for $R(s)$.  Dubnicka followed
Marshall in his 1992 preprint \cite{DDS92}, and in the present calculation
both Marshall's fitted parameterization (with his errors) and trapezoidal
integration \cite{EJ95} have been compared for the higher-energy contributions
to \hbox{$a_{\mu}^{had}$}. 
As the QCD and data-based results are in good agreement
it is clear that the second order QCD expression is sufficient.

\section{WFSA Results and Projections}
\label{s:proj}

\subsection{Summary Results}

As the preceding discussion of the WFSA evaluation shows,
the core \hbox{$a_{\mu}^{had}$}\ problem is how to combine the correlations
among the systematic errors. 
In the present calculation,
partial \hbox{$a_{\mu}^{had}$}\ integrations have been performed
first so relative errors and
error-weighted
averaged systematic and statistical 
errors could be defined
at an early stage of the error combination
calculation. The averaged systematic errors have been
correlated according to the postulates discussed in Sect.~\ref{s:post}
for all experiments measuring more than one
hadronic mode; all other errors have been combined in quadrature
only after such correlations have been taken into account.

All of the WFSA calculations have been collected in Tab.~\ref{t:grand}.
The first two lines of the table are summary results
of the \hbox{$a_{\mu}^{had}$}\ contributions and errors presented in 
Table~\ref{t:upto2}. In the $>$ 2 hadrons line of Tab.~\ref{t:upto2}
the results from Tab.~\ref{t:hmm} are presented.
The third line line in Tab.~\ref{t:grand} for 
$R(s)$ from 2.0 - 3.1 GeV is the summary result
from Table~\ref{t:Rwf}. All of the $a_{\mu}^{had}$ central values
are averaged over detectors first and then summed over modes,
while the errors are combined over modes first (to correlate
systematics) and only combined over detectors in the end.
The errors in Tab.~\ref{t:grand} may be combined in quadrature
since there are no correlations remaining among the categories chosen.
(This feature is not completely 
present in the previous calculations of $a_{\mu}^{had}$.)

A subtotal for $a_{\mu}^{had}$ contributions below 3.1 GeV
is presented in Tab.~\ref{t:grand} in order to show the total errors
before the use of QCD. It is there seen that
the QCD results 
do not have much influence on the final results
since the error from the lower energy ($<$ 3.1 GeV) region dominates. 
For comparison, if in place of the QCD calculation
the data based evaluation for the region above 3.1 GeV \cite{EJ95} is used
the final errors are only slightly higher:
$a_{\mu}^{had} = 702.718(7.787)(14.147)$.
Hence,
the WFSA result is thus shown to be largely 
insensitive to the use of QCD 
or the experimental data after 3.1 GeV.
(This justifies both the use of QCD down to the energy 3.1 GeV and
the neglect of third order QCD terms; as new data bring the errors
from the lower energy regions downward (see Sect.~\ref{s:proj}), 
the QCD error will become more 
significant however.)

It is further apparent from Table~\ref{t:grand} that the error in the energy
region below 1.4 GeV dominates the total error
in \hbox{$a_{\mu}^{had}$}, and this error comes 
mostly from the Olya \hbox{$\pi^+ \pi^-$}\
data. (One advantage of the WFSA method is 
that in presenting averaged errors (both statistical and systematic
separately)
for each detector and exclusive mode, it remains possible to
identify which detector and exclusive mode is contributing the most to the
overall total error.)
The WFSA method most heavily weights
the data with the smallest errors (see Table~\ref{t:systpp}),
and hence most of the stated systematic
errors are indeed being added linearly for the experiments with the
smallest errors. 

\subsection{Energy Region $<$ 1.4 GeV}

The overall WFSA results indicate that 
improved measurements below 1.4 GeV, 
can 
significantly 
improve 
upon the overall lowest order
hadronic vacuum polarization uncertainty in $a_{\mu}^{had}$.
To show this, the WFSA procedure has been performed with an additional two 
experiments in view, the CMD2 experiment at VEPP-2M in Novosibirsk, Russia
and a hypothetical detector at DAFNE at Frascati, Italy. The 
$a_{\mu}^{had}$
results for the energy region below 1.4 GeV
are shown in Tab.~\ref{t:newdata} for three cases:
a) without new data, b) with expected 
errors (stat)(sys) = (3\%)(0.5\%) from the CMD2 experiment
and c) CMD2 plus a second experiment with errors (stat)(sys) = (3\%)(0.5\%).
In particular, the central values for hadronic cross sections
have been chosen equal to CMD or Olya values
(and hence the $S$ factors will generally be less than one and so not
applied),
while the errors above
have been 
determined 
for only the following exclusive hadronic modes:
$\pi^+ \pi^-$ (2 $\pi$), $\pi^+ \pi^- \pi^0$ (3 $\pi$),
$\pi^+ \pi^- \pi^+ \pi^-$ (4 $\pi$), $\pi^+ \pi^- \pi^0 \pi^0$ (4 $\pi$),
$\pi^+ \pi^- \pi^+ \pi^- \pi^0$ (5 $\pi$), $K^+ K^-$ and $K_L K_S$.

The choice of 3 \% statistical error is
based on the 
CMD2 data taking assumption of 1000 pion pairs per energy point
since $\pi^+ \pi^-$ is the dominant contribution to $a_{\mu}^{had}$.
Since the luminosity collected by CMD2 has been determined by this requirement
on the $\pi^+ \pi^-$ cross section (and by competing requirements
for other vector meson physics goals),
the actual statistical errors on the non-$\pi^+ \pi^-$ higher multiplicity
modes will in fact be somewhat different, but this is a higher order effect
here neglected.

The choice of 0.5 \% systematic error is based on the fact that the limiting
error on the new CMD2 cross section measurements appears to be the
error on higher order corrections to the QED Bhabha cross sections
used in calculations of the luminosity.
While radiative
corrections have been calculated to good accuracy (0.11 \% \cite{LEP})
for the $t$-channel contributions
to Bhabha scattering
useful for the forward region luminosity monitors in use at LEP,
they are not useful for CMD2; the principle luminosity
is determined there by large angle Bhabha events in the barrel calorimeter
where significant $t$- and $s$- channel interference terms are present.
The task of calculating these interference terms to 0.5 \% accuracy is
well underway \cite{BINP96} and it is assumed that this will be the limiting
error on the new CMD2 cross section measurements.

The results in Tab.~\ref{t:newdata} show that without new data
the errors on the contributions to
$10^{10} a_{\mu}^{had}$ 
below 1.4 GeV
will be 
$10^{10}$ (stat)(sys) = (7.4)(13.1)
and with particular new data they can become 
$10^{10}$ (stat)(sys) = (2.1)(2.9)
which is equivalent to a total error of $3.6 \times 10^{-10}$.
Comparing this with the 
AGS E821 experimental error goal of 
$\sigma_{exp}^{a_{\mu}} \sim \pm \ 4.0 \times 10^{-10}$
it is clear that the new data is needed. In particular, if this data
has errors of 3 \% statistical and 0.5 \% systematic
on the stated modes, then they will in fact be sufficient to reduce
the hadronic contribution errors in the energy range below 1.4 GeV
to below the error goal of the new measurement of
$a_{\mu}^{had}$ by the BNL AGS E821 Collaboration.

\subsection{Energy Region $>$ 1.4 GeV}

However, the new data from BINP Novosibirsk and INFN Frascati
will not reduce the errors on $a_{\mu}^{had}$ in
the energy range above 1.4 GeV. 
The current error on 
$a_{\mu}^{had}$
obtained by use of the WFSA procedure in the region above 1.4 GeV
is $10^{10}$ (stat)(sys) = (2.4)(5.3).
In this region,
most of the error comes from the \hbox{$\gamma \gamma 2$} 
detector at the former 3 GeV Adone storage ring in INFN Frascati, Italy.
As this error is larger than the
BNL AGS E821 experimental error goal of 
$\sigma_{exp}^{a_{\mu}} \sim \pm \ 4.0 \times 10^{-10}$,
clearly more experiments are needed for interpretation
of the new measurement of $a_{\mu}^{had}$. Fortunately,
it is possible to use new high statistics data on $\tau$-decays
for new measurements of multi-pion production in the energy
region 1.4 - 2.0 GeV, and above 2.2 GeV there are plans to make
further measurements of the cross section ratio
$R(s)$ at BEPC Beijing, China \cite{BLR96}.

\subsubsection{$\tau$ Decay Data for Energy Region 1.4 - 2.0 GeV}

The new high statistics data on $\tau$-decays 
has already been included in the present calculation by use of the
$\omega \pi^0$ cross section obtained from the ARGUS Collaboration 
\cite{WP_ARGUS} \cite{ND91}.
The idea is based on the fact that the coupling of $W^{\pm}$ 
bosons to quarks is related to
the photon coupling to quarks by an isospin rotation (CVC relation)
\cite{GR85}. The effect on the total errors in the energy region
1.4 - 2.0 GeV 
are presented in Tab.~\ref{t:newdata14} where the assumed errors
are (stat)(sys) = (3\%)(1\%) for the following hadronic modes:
$\pi^+ \pi^- \pi^+ \pi^-$ (4 $\pi$ charged),
$\pi^+ \pi^- \pi^0 \pi^0$ (4 $\pi$ neutral),
$\pi^+ \pi^- \pi^+ \pi^- \pi^0$ (5 $\pi$) and
$\pi^+ \pi^- \pi^+ \pi^- \pi^0 \pi^0$ (6 $\pi$).

New data on the 4 $\pi$ neutral, 5 $\pi$ and 6 $\pi$ modes can 
reduce the
$\gamma \gamma 2$ contribution to the total errors (as shown in Row 2)
while new data on the 4 $\pi$ charged mode is required to 
reduce the DM1 and DM2 contribution to the total errors (as shown in Row 3).
Therefore,
if many data points ($\sim 10 - 20$) for each of the specified 
modes across the energy
region 1.4 to 2.0 GeV
with the assumed errors can be extracted from $\tau$-decay
data,
then the contribution to the total errors from this energy
region will be reduced significantly as shown in Tab.~\ref{t:newdata14}.

\subsubsection{New $R(s)$ Measurements for Energy Region 2.0 - 3.1 GeV}

The effect of new data on the error in the 2.0 - 3.1 GeV energy region
is presented in Tab.~\ref{t:newdata23}.
The WFSA procedure has been performed with a new
hypothetical detector measuring $R(s)$ from 2.2 - 3.1 GeV
(central value of $R(s) = 3.0$ for all points so no $S$-factors exceed 1) 
with assumed errors of (stat)(sys) = (3\%)(0.5\%).
If new $R(s)$ data in this energy region can be obtained with these
errors then the $\gamma \gamma 2$ contribution to the total error 
can be significantly reduced.

\subsection{Conclusion}

Taking the new hypothetical data together 
(including $\tau$ decays and $R(s)$ measurements),
the total error from the energy
range above 1.4 GeV (including resonances and QCD) then becomes 
$10^{10}$ (stat)(sys) = (0.6)(1.6). 
Combining the hypothetical results from above and below 1.4 GeV the total error
becomes $10^{10}$ (stat) (sys) = (2.2)(3.3) which is equivalent to a total
error of $3.98 \times 10^{-10}$. Therefore, new measurements of $e^+e^- \to
hadrons$ (including the cross section ratio $R(s)$) and $\tau$ decays with
assumed errors of (stat)(sys) = (3\%)(0.5\%) and (3\%)(1.0\%) respectively are
sufficient to reduce the total error on the lowest order contribution to
$a_{\mu}^{had}$ in all energy regions below the expected error of the new
measurement of $a_{\mu}$ by the AGS E821 Collaboration. 

\bigskip\noindent
{\bf Acknowledgments}
\bigskip

The authors would like to acknowledge useful discussions with 
S.I. Eidelman, I. Grosse, 
T. Kinoshita, 
W.J. Marciano and V.G. Zavarzin in various aspects of this work. 
We would also
like to thank B.I. Khazin, E.P. Solodov and V.G. Zavarzin 
for discussions about recent and expected
CMD2 results. Comments on previous drafts by R.M. Carey, J.P. Miller
and B.L. Roberts are also appreciated. 
Finally, a very special thanks is extended to V.A. Monich
for help improving the WFSA calculation routines.

  \appendix
\section{Experimental Data Used}
\label{a:data}

The experimental data we used for our evaluation of $a_{\mu}^{had}$
is listed by hadronic exclusive mode in 
Tab.\ \ref{t:data}.


\end{multicols}

\newpage

\widetext

\begin{table}
\caption{History of $a_{\mu}^{had}$ calculations.
The values 4 and $15 \times 10^{-10}$ 
are the AGS E821 error goal and electroweak contribution to $a_{\mu}$
respectively.}
\begin{tabular}{|l|c|c|c|} \hline
\multicolumn{4}{|c|}{Calculations based on Experimental Data}\\ \hline
Author(s)                 & Year & 
$10^{10} a^{had}_{\mu} (stat) (sys) $ & ref \\
\hline
Budker Institute & 1985 & 684(11)        & \cite{BINP85} \\
Kinoshita, Nizic, Okamoto 
                 & 1985 & 707(6)(17)     & \cite{KNO85}  \\ 
Dubnickova, Dubnicka, Stricinec 
                 & 1992 & 699(4)(2)$^a$  & \cite{DDS92}  \\
Eidelman, Jegerlehner 
                 & 1995 & 702(6)(14)     & \cite{EJ95}   \\
Apel, Yndurain
                 & 1995 & 710(11)        & \cite{AY95}  \\
WFSA evaluation       
                 & 1996 & 703(8)(14)     &               \\ \hline
\end{tabular}
\smallskip
$^a$ Improvement due mostly to fitting function technique. See text.

\label{t:history}
\end{table}

\begin{table}
\caption{The Grand Total WFSA $10^{10} a_{\mu}^{had}$ results}
\label{t:grand} 
\begin{tabular}{|c||c|} \hline
Energy Region [GeV]  & WFSA $10^{10} a_{\mu}^{had}$ (stat) (sys)\\ \hline \hline
$\sigma(e^+e^- \to hadrons)$ $E_{\theta}$ - 1.4 & 611.332 (7.399) (13.045) \\
$\sigma(e^+e^- \to hadrons)$          1.4 - 2.0 &  32.466 (0.756) ( 2.379) \\
R(s) 2.0 - 3.1                        & 28.374 (2.288) (4.400)$^{a}$ \\
$J/\Psi$ (6 states)                   &  9.047   (-)   (0.969)      \\
$\Upsilon$ (6 states)                 &  0.109   (-)   (0.013)      \\ 
QCD 3.1 - $\infty$                    &  21.301  (-)  (0.371)$^{b}$ \\ 
\hline \hline
Sub Total $<$ 3.1 + $J/\Psi,\Upsilon$ & 681.328 (7.782) (14.005) \\ \hline
Sub Total $<$ 1.4                     & 611.332 (7.399) (13.045) \\ \hline
Sub Total $>$ 1.4                     &  91.297 (2.410) ( 5.108) \\ 
\hline \hline 
Total Figure~\ref{f:had1}             & 702.629 (7.782) (14.009) \\ \hline
\end{tabular}

$E_{\theta}$ refers to the particular thresholds of the 
exclusive hadronic modes.

$^{a}$ Represents systematic errors mainly from $\gamma \gamma 2$ detector 
added linearly for energy regions 1.4 - 2.0 and 2.0 - 3.1 GeV (1.7 + 2.7).

$^{b}$ Errors determined by 
$a_{\mu}^{QCD} 
\left( \Lambda_{\overline{MS}} \pm \Delta \Lambda_{\overline{MS}} \right)$
\end{table}

\begin{table}
\vbox{
\caption{Our WFSA $10^{10} a_{\mu}^{had}$ results below 2.0 GeV. Upper (lower)
contributions listed
in the total column are for below (above) 1.4 GeV. Upper (lower)
numbers in parentheses are contributions to the
statistical (systematic) errors.
Systematic (statistical) errors are combined linearly (in quadrature)
in each column separately.}
\smallskip
\label{t:upto2} 
\begin{tabular}{|c||c|c|c|c||c|c|c||c|} \hline
Mode     & CMD     & Olya    & ND & DM1A & DM1D     & DM2 & Other & Total \\ 
\hline \hline
         &    -    &    -    &    &    -    & &    -    &    -    & 502.184\\ 
$\pi^+\pi^-$
         & (4.806) & (1.306) &    & (5.064) & & (0.058) & (0.814)  &   0.781 \\ 
         & (4.789) & (5.807) &    & (4.021) & & (0.110) & (0.732)  &        \\ 
\hline
         &    -    & &    -    &    -    & &    -    &    -    &  50.886 \\ 
$\pi^+\pi^-\pi^0$
         & (0.509) & & (0.856) & (1.209) & & (0.041) & (0.049) &   0.677 \\ 
         & (0.104) & & (1.303) & (0.715) & & (0.053) & (0.009) &        \\ 
\hline
               &   -   &   -   &   -   & &   -   &   -   &   -   & 16.782 \\
$>2$ hadrons   &(0.236)&(0.369)&(0.205)& &(0.241)&(0.307)&(0.631)& 29.995 \\
               &(0.589)&(2.405)&(2.070)& &(1.071)&(1.857)&(1.735)&       \\ 
\hline
         &    -    &    -    & & & &    -    &                  & 20.623 \\ 
$K^+K^-$
         & (0.119) & (0.953) & & & & (0.046) &                  & 4.45$^{a}$ \\ 
         & (0.058) & (1.900) & & & & (0.090) & (0.230)$^{b}$ & 0.759 \\ 
\hline
         &    -    &    -    & & &    -    & & &  0.755       \\ 
$K_L K_S$
         & (0.173) & (0.156) & & & (0.033) & & & 14.07$^{a}$   \\
         & (0.040) & (0.101) & & & (0.017) & & (0.586)$^{b}$ & 0.154 \\ 
\hline
         & & & & &    -    & &         &        \\ 
$p \overline{p}$
         & & & & & (0.020) & &         &  0.100  \\ 
         & & & & & (0.010) & &         &        \\ 
\hline
$\omega \to \pi^0 \gamma, \eta \gamma$ 
                  & & & &&       &       & (0.040)$^{b}$ &   0.980$^{a}$ \\ 
\hline
$\phi \to \pi^0 \gamma, \eta \gamma$ 
                  & & & &&       &       & (0.024)$^{b}$ &   0.602$^{a}$ \\
\hline \hline
                &   -   &    -    &   -   &       & & &       & 611.33   \\
Total           &(4.842)& (1.666) &(0.880)&(5.206)& & & (0.807) & (7.399)  \\
$E_{\theta}$ - 1.4  
                &(5.580)& (10.213) &(3.373)&(4.736)& & & (0.962) & (13.045) \\
\hline \hline
                & & & & &   -   &   -   &   -           & 32.47   \\
Total           & & & & &(0.244)&(0.318)& (0.642)       & (0.756) \\
1.4 - 2.0       & & & & &(1.098)&(2.110)& (1.747)$^{c}$ & (2.379) \\
\hline
\end{tabular} 
$E_{\theta}$ refers to the particular thresholds of the
exclusive hadronic modes. 

DM1A = DM1 at ACO; DM1D = DM1 at DCI, Orsay.

$^{a}$ Integration of energy dependent width Breit-Wigner in absence of data.

$^{b}$ Errors determined by $m_V$, $\Gamma_V$, $B_{V \to ee}$ derivatives
of the Breit-Wigner.

$^{c}$ The $\gamma \gamma 2$ systematic errors from energy regions
1.4 - 2.0 GeV (1.7) and 2.0 - 3.1 GeV (2.7) will be added linearly
and presented in Table~\ref{t:grand}.
}
\end{table}

\begin{table}[H]
\caption{Our WFSA $10^{10} a_{\mu}^{had}$ results
for $>$ 2 hadrons exclusive modes
below 2.0 GeV. Upper (lower) contributions listed
in the total column are for below (above) 1.4 GeV. Upper (lower)
numbers in parentheses are contributions to the
statistical (systematic) errors.
Systematic (statistical) errors are combined linearly (in quadrature)
in each column separately.}
\smallskip
\label{t:hmm} 
\begin{tabular}{|c||c|c|c||c|c|c||c|c|} \hline
Mode     & CMD     & Olya    & ND & DM1D  & DM2 & Other & Total & S Factor \\ 
\hline \hline
$\pi^+\pi^-\pi^0\pi^0$
         && (0.325) & (0.193) & & (0.256) & (0.335) & 10.783 & 1.614 \\ 
         && (1.491) & (1.514) & & (0.868) & (0.841) &  8.543 & 1.692 \\ 
\hline
$\pi^+\pi^-\pi^+\pi^-$
        & (0.225) & (0.174) & (0.066) & (0.104) & (0.070) & &  5.161 & 0.486 \\ 
        & (0.527) & (0.914) & (0.526) & (0.623) & (0.628) & & 10.215 & 1.221 \\ 
\hline
$\pi^+\pi^-\pi^+\pi^-\pi^0\pi^0$
         & & & &         &         & (0.494) &             & \\ 
         & & & &         &         & (0.763) & 5.089       & 0.0 \\ 
\hline
$A_1 (\pi^+\pi^-\pi^+\pi^-\pi^0)$
         & (0.072) & & & (0.124) & & (0.187) & 0.305       & 0.0   \\ 
         & (0.062) & & & (0.192) & & (0.131) & 2.533       & 1.010 \\ 
\hline
$K_S K^{\pm} \pi^{\mp} $
         & & & & (0.158) & (0.104) &         &  & \\ 
         & & & & (0.0)   & (0.119) &         &  0.951         & 3.199 \\ 
\hline
$K^+K^-\pi^+\pi^-$
         & & & & (0.072) & (0.091) &         &  & \\ 
         & & & & (0.135) & (0.123) &         &  0.815         & 2.709 \\ 
\hline
$K^* K^{\pm} \pi^{\mp} $
         & & & &         & (0.057) &         &  & \\ 
         & & & &         & (0.069) &         &  0.692         & 0.0 \\ 
\hline
$B_1(\omega \pi^0)$ 
           & & & (0.017) & &         & (0.079) & 0.533  & 1.132 \\ 
           & & & (0.030) & &         & (0.021) & 0.210  & 0.0 \\ 
\hline
$B_2(\eta \pi^+\pi^-)$ 
            & & & &         & (0.039) &         &  & \\ 
            & & & &         & (0.044) &         &  0.444 & 0.0 \\ 
\hline
$\pi^+\pi^-\pi^+\pi^-\pi^+\pi^-)$ 
         & & & & (0.040) &         &         &  & \\ 
         & & & & (0.117) &         &         &  0.419 & 0.0 \\ 
\hline
$B_3(\omega \pi^+\pi^-)$ 
             & & & & (0.005) & (0.003) &         &  & \\
             & & & & (0.003) & (0.004) &         &  0.084 & 0.982 \\ 
\hline \hline
       &   -   &   -   &   -   &   -   &   -   &   -   & 16.782  &         \\
Total  &(0.236)&(0.369)&(0.205)&(0.241)&(0.307)&(0.631)& 29.995  & \\
       &(0.589)&(2.405)&(2.070)&(1.071)&(1.857)&(1.735)&    -    & \\ \hline
\end{tabular} 
\end{table}

\begin{table}[H]
\caption{Our WFSA $10^{10} a_{\mu}^{had}$ results 
for $R(s)$ from 2.0 - 3.1 GeV.
Error weighted averages with $S$-factor in ``Total" column. Numbers in 
upper (lower) parenthesis are WFSA contributions to the 
statistical (systematic) errors which 
are combined in quadrature (linearly) in each column separately.
Lower right corner errors are quadrature sums of errors in the ``Total"
row.}
\label{t:Rwf} 
\begin{tabular}{|c||c|c|c||c|c|} \hline
Energy Range [GÅV]
           & BCF     & $\gamma \gamma 2$ & Mark I  & Total   & S Factor \\ 
\hline \hline
           & 18.778  & 22.441  &         & 20.480  & 0.637 \\ 
2.0 - 2.6  & (2.149) & (0.527) &         & & \\ 
           & (0.207) & (2.093) &         & & \\ 
\hline
           &  3.718  &  6.535  &  4.781  &  4.898  & 1.210 \\ 
2.6 - 2.87 & (0.469) & (0.130) & (0.236) & & \\ 
           & (0.032) & (0.451) & (0.406) & & \\ 
\hline
           &  1.721  &         &  1.777  &  1.742  & 0.096 \\ 
2.87 - 3.0 & (0.202) &         & (0.061) & & \\ 
           & (0.019) &         & (0.193) & & \\ 
\hline
           &         &         &  1.254  &  1.253  & 0.0   \\ 
3.0 - 3.1  &         &         & (0.062) & & \\ 
           &         &         & (0.313) & & \\ 
\hline \hline
           &    -    &    -    &    -    & 28.374  & \\
Total      & (2.208) & (0.543) & (0.251) & (2.288) & \\
           & (0.258) & (2.544) & (0.913) & (2.700) & \\ \hline
\end{tabular} 
\end{table}

\begin{table} 
\caption{WFSA \hbox{$10^{10} a_{\mu}^{had}$} detector total statistical (upper)
and systematic (lower) errors 
below 1.4 GeV: Row 1) with no new data, Row 2) one new experiment 
with 3 \% stat., 0.5 \% syst. errors, Row 3) two new experiments 
with 3 \% stat. and 0.5 \% syst. errors.}
\label{t:newdata} 
\begin{center}
\begin{tabular}{|c|c|c|c|c|c|c||c|} \hline
\multicolumn{8}{|c|}{WFSA $10^{10} a_{\mu}^{had}$ Errors Below 1.4 GeV} 
\\ \hline
New 2  & New 1 & CMD   & Olya   & ND    & DM1A  & Other & Total   \\ \hline 
       &       &       &        &       &       &       &         \\  
       &       &(4.842)& (1.666)&(0.880)&(5.206)&(0.807)& (7.399) \\  
       &       &(5.580)&(10.213)&(3.373)&(4.736)&(0.962)&(13.045) \\  \hline
       &       &       &        &       &       &       &         \\  
       &(2.426)&(1.266)& (1.148)&(0.136)&(1.239)&(0.451)& (3.250) \\ 
       &(2.224)&(1.292)& (3.608)&(0.405)&(1.241)&(0.760)& (4.681) \\  \hline
       &       &       &        &       &       &       &         \\  
(1.183)&(1.183)&(0.789)& (0.528)&(0.078)&(0.766)&(0.361)& (2.103) \\  
(1.256)&(1.256)&(0.773)& (1.921)&(0.244)&(0.769)&(0.718)& (2.934) \\  \hline
\end{tabular} 
\end{center}
\end{table} 

\begin{table} 
\caption{WFSA \hbox{$10^{10} a_{\mu}^{had}$} detector total statistical (upper)
and systematic (lower) errors in region 1.4 - 2.0 GeV.
Row 1) with no new data, Row 2) new hypothetical data (possibly
from $\tau$ decay spectral functions) on modes 
$\pi^+\pi^-\pi^0\pi^0$,
$\pi^+\pi^-\pi^+\pi^-\pi^0$ and
$\pi^+\pi^-\pi^+\pi^-\pi^0\pi^0$ 
with 3 \% stat., 1.0 \% syst. errors, Row 3) including as well the mode
$\pi^+\pi^-\pi^+\pi^-$.}
\label{t:newdata14} 
\begin{center}
\begin{tabular}{|c|c|c|c||c|} \hline
\multicolumn{5}{|c|}{WFSA $10^{10} a_{\mu}^{had}$ Errors in 1.4 - 2.0 GeV 
Region}
\\ \hline
New 1  & DM1D  & DM2   & $\gamma \gamma 2$ & Total \\ \hline 
       &       &       &                   &\\  
       &(0.244)&(0.318)&(0.628)            &(0.745)\\
       &(1.098)&(2.110)&(1.744)            &(2.950)\\  \hline
       &       &       &                   &\\  
(0.068)&(0.210)&(0.192)&(0.067)            &(0.300)\\
(0.115)&(0.922)&(1.320)&(0.141)            &(1.620)\\  \hline
       &       &       &                   &\\  
(0.082)&(0.179)&(0.157)&(0.067)            &(0.261)\\
(0.175)&(0.346)&(0.514)&(0.141)            &(0.659)\\  \hline
\end{tabular} 
\end{center}
\end{table}

\begin{table} 
\caption{WFSA \hbox{$10^{10} a_{\mu}^{had}$} detector total statistical (upper)
and systematic (lower) errors in region 2.0 - 3.1 GeV.
Row 1) with no new data, Row 2) one new experiment above 2.2 GeV
with 3 \% stat., 0.5 \% syst. errors}
\label{t:newdata23} 
\begin{center}
\begin{tabular}{|c|c|c|c||c|} \hline
\multicolumn{5}{|c|}{WFSA $10^{10} a_{\mu}^{had}$ Errors in 2.0 - 3.1 GeV 
Region}
\\ \hline
New 1  & BCF   & $\gamma \gamma 2$ & Mark I & Total \\ \hline 
       &       &                   &        &\\  
       &(2.208)&(0.543)            &(0.251) &(2.288)\\ 
       &(0.258)&(2.544)            &(0.913) &(2.715)\\  \hline
       &       &                   &        &\\  
(0.109)&(0.485)&(0.140)            &(0.025) &(0.517)\\  
(0.058)&(0.109)&(0.601)            &(0.084) &(0.619)\\  \hline
\end{tabular} 
\end{center}
\end{table}

\begin{table} 
\begin{center}
\caption{Systematic Errors used in our evaluation 
of the $\pi^+ \pi^-$ contribution to 
\hbox{$a_{\mu}^{had}$}.}
\label{t:systpp} 
\begin{tabular}{|c|c|} \hline
Detector  & Systematic error \\ \hline \hline
NA7       & .01,.02,.025,.05 \\
CMD       & .02            \\
TOF       & .035$^a$  \\
DM1       & .022           \\
Olya      & .04 - .15      \\
M2N       & .0$^b$    \\
BCF       & .1$^c$         \\
$\mu \pi$ & .1$^c$         \\
MEA       & .1$^c$         \\
DM2       & .12            \\ \hline
\end{tabular} 
\end{center}

\bigskip

$^a$ evaluated from errors on radiative correction/efficiency factors.

$^b$ systematic error is included with statistical in quoted errors.

$^c$ Arbitrary value - inconsequential since statistical errors are large.
\end{table}

\begin{table} 
\caption{Systematic Errors used in our evaluation of 
the $\pi^+ \pi^- \pi^0$ contribution to 
\hbox{$a_{\mu}^{had}$}.}
\label{t:systppp} 
\begin{center}
\begin{tabular}{|c|c|c|} \hline 
Detector & Energy Range [GeV] & Systematic error \\ \hline \hline
CMD  &               .76  -  .81  & .0$^a$     \\
CMD  &               .84  - 1.013 & .07    \\
ND   &               .414 -  .765 & .1     \\
ND   &               .805 - 1.003 & .1     \\
ND   &              1.036 - 1.379 & .2     \\
DM1  &              0.414 - 1.098 & .032   \\
DM2  &              1.34  - 2.0   & .0866  \\
$\gamma \gamma 2$ & 1.437 - 2.0   & .15    \\ \hline
\end{tabular} 
\end{center}


$^a$ systematic errors are less than quoted statistical errors.
\end{table}

\begin{table} 
\caption{Systematic errors used in our evaluation of higher multiplicity
exclusive hadronic mode \hbox{$a_{\mu}^{had}$}\ contribution.}
\label{t:systothr} 
\begin{center}
\begin{tabular}{|c||c|c|c|c||c|c|c|} \hline
Mode   & Olya   & CMD   & ND & ARGUS & DM1     & DM2  & $\gamma \gamma 2$ \\ 
\hline \hline
\hbox{$K^+ K^-$}	&0.1$^a$ & 0.038 & 0.1  &      &         & 0.1$^a$ & \\
\hbox{$K_L K_S$}	&0.3     & 0.1   & 0.1  &      & 0.1$^a$ &         & \\
\hbox{$\pi^+ \pi^- \pi^+ \pi^-$}	
        &0.15    & 0.1   & 0.1  &      & 0.1     & 0.1$^a$ & \\
\hbox{$\pi^+ \pi^- \pi^0 \pi^0$}
        &0.15    &       & 0.15 &      &         & 0.1     & 0.15 \\
\hbox{$\pi^+ \pi^- \pi^+ \pi^- \pi^0$}	
        &        & 0.20  &      &      & 0.12    &         & 0.15 \\
\hbox{$K^+ K^- \pi^+ \pi^-$}	
        &        &       &      &      &  0.13   & 0.1$^a$ & \\ 
\hbox{$K_S K^{\pm} \pi^{\mp}$}	
        &        &       &      &      &   0.0   & 0.1$^a$ & \\
\hbox{$K^* K^{\pm} \pi^{\mp}$}	
        &        &       &      &      &         & 0.1$^a$ & \\
\hbox{$\pi^+ \pi^- \pi^+ \pi^- \pi^+ \pi^-$}	
                  &        &       &      &      & 0.28    &         & \\
\hbox{$\pi^+ \pi^- \pi^+ \pi^- \pi^0 \pi^0$}	
                  &        &       &      &      &         &         & 0.15 \\
\hbox{$p \overline{p}$}	&        &       &      &      & 0.1$^a$ &         & \\ 
$\omega \pi$      &        &       & 0.1  & 0.1  &         &         & \\ 
$\omega \pi \pi$  &        &       &      &      & 0.12    & 0.082   & \\ 
$\eta \pi \pi$	  &        &       &      &      &         & 0.1$^a$ & \\ \hline
\end{tabular} 
\end{center}

\bigskip

$^a$ systematic error not discussed in reference
\end{table}

\begin{table} 
\caption{Systematic errors used in evaluation of the energy region
2.0 - 3.1 GeV
\hbox{$a_{\mu}^{had}$}\ contribution.}
\label{t:syst23} 
\begin{center}
\begin{tabular}{|c||c|c|c|} \hline
Energy Region [GeV]
          & BCF   & $\gamma \gamma 2$   & Mark I \\ \hline \hline
2.0 - 3.1 & 0.02  & 0.21   & 0.25   \\ \hline
\end{tabular} 
\end{center}
\end{table}

\begin{table}
\caption{Data used in the WFSA evaluation}
\label{t:data}
\begin{tabular}{|c|l|l|l|l|l|l|} \hline
Mode & {OLYA} & {CMD} & {ND} & {DM1} & {DM2} & 
Others \\ \hline
\hbox{$\pi^+ \pi^-$} 
& & & & & & 
{M2N}  {BCF}  {$\mu\pi$}  {MEA} {NA7}  {TOF}  
 \\ &
\cite{2PI_OLYA} &
\cite{BINP85}   & &
\cite{2PI_DM1} &
\cite{2PI_DM2} &
\cite{2PI_M2N} \ \ \ \ 
\cite{2PI_BCF} \ \ \ \ 
\cite{2PI_MUPI} \ \ \ \ 
\cite{2PI_MEA} \ \ \ \ 
\cite{2PI_NA7} \ \ \ \ 
\cite{2PI_TOF} \\ \hline

\hbox{$\pi^+ \pi^- \pi^0$} 
&  & & & & & {GG281} \\
&  & 
\cite{3PI_CMD87} \cite{3PI_CMD89} &
\cite{ND91} &
\cite{3PI_DM1} &
\cite{3PI_DM2} &
\cite{GG281} \\ \hline

\hbox{$K^+ K^-$} 
& & & & & & {MEA} \\ 
& 
\cite{OLYA82} &
\cite{CMD83} &
& &
\cite{KK_DM2} & \cite{2PI_MEA} \\ \hline

\hbox{$K_L K_S$}  
& & & & & &  \\ 
& 
\cite{OLYA82} &
\cite{CMD83} &
&
\cite{KKZ_DM1} & & \\ \hline

$\pi^+ \pi^- \pi^+ \pi^-$  
& & & & & &  \\ 
& 
\cite{4PI_OLYA} &
\cite{4PI_CMD} &
\cite{ND91} &
\cite{4PI_DM1} &
\cite{DM290} & \\ \hline

$\pi^+ \pi^- \pi^0 \pi^0$  
& & & & & & {GG281}  \\ 
& 
\cite{4PIZ_OLYA} & &
\cite{ND91} & & 
\cite{KK_DM2} & 
\cite{GG281} \\ \hline

$\pi^+ \pi^- \pi^+ \pi^- \pi^0$  
&  & & & & & {GG281} \\
&  &
\cite{CMD83} & &
\cite{5PI_DM1} & &
\cite{GG281} \\ \hline

$\pi^+ \pi^- \pi^+ \pi^- \pi^+ \pi^-$  
& & & &  & & \\
& & & &  \cite{6PI_DM1} & & \\ \hline

$\pi^+ \pi^- \pi^+ \pi^- \pi^0 \pi^0$  
& & & & & &  {GG281} \\
& & & & & &  \cite{GG281} \\ \hline

\hbox{$K_S K^{\pm} \pi^{\mp}$}  
& & & & & &\\
& & & & \cite{KKP_DM1} & \cite{DM290} &\\ \hline

\hbox{$K^+ K^- \pi^+ \pi^-$}  
& & & &  & & \\
& & & &  \cite{KKPP_DM1} & \cite{DM290} & \\ \hline

\hbox{$K^* K^{\pm} \pi^{\mp}$}  
& & & & &  & \\
& & & & &  \cite{DM290} & \\ \hline

$\omega \pi^0$  
& & &  & & & {ARGUS} \\
& & &  \cite{ND91} & & & \cite{WP_ARGUS} \\ \hline

$\eta \pi^+ \pi^-$  
& & & & &  & \\
& & & & &  \cite{EPP_DM2} & \\ \hline

$\omega \pi \pi$  
& & & &  & & \\
& & & &  \cite{WPP_DM1} & \cite{WPP_DM2} & \\ \hline

\hbox{$p \overline{p}$}  
& & & &  & & \\
& & & &  \cite{PPB_DM1} & \cite{PPB_DM2} & \\ \hline

$R(s)$ 2.0 - 3.1 GeV  
& & & & & &  {GG279}  {GG281}  {BCF74}  {MarkI} \\
& & & & & &  
\cite{GG279}  \ \ \ \ \ 
\cite{GG281}  \ \ \ \ \ 
\cite{BCF74}  \ \ \ \ \ 
\cite{MarkI}
\\ \hline
\end{tabular}
\end{table}

\end{document}